\documentclass[twocolumn,superscriptaddress,aps,prb]{revtex4-2}
\usepackage{graphicx}
\usepackage{dcolumn}
\usepackage{bm}
\usepackage{subfigure}
\usepackage{amsmath}
\usepackage{mathrsfs}
\usepackage{amssymb}
\usepackage{setspace}
\usepackage{array}
\usepackage{multirow}
\usepackage{float}
\usepackage{flushend}
\usepackage{footmisc}
\usepackage[pdfstartview=FitH,
CJKbookmarks=true,
colorlinks,
linkcolor=blue,
anchorcolor=blue,
citecolor=blue,
urlcolor=blue,
]{hyperref}

\usepackage{footmisc}

\renewcommand{\footnoterule}{
	\kern -4pt  
	\hrule width 0.18\linewidth height 0.6pt
	\kern 12pt 
}

\usepackage{footmisc}

\usepackage{braket}
\usepackage{braket}
\begin{document}

\global\long\def\id{\mathbbm{1}}
\global\long\def\ui{\mathbbm{i}}
\global\long\def\ud{\mathrm{d}}

\title{Expedited thermalization dynamics in incommensurate systems}
\author{Mingdi Xu}
\affiliation{School of Physics, Nankai University, Tianjin 300071, China}
\author{Zijun Wei}
\affiliation{School of Physics, Nankai University, Tianjin 300071, China}

\author{Xiang-Ping Jiang}
\email{2015iopjxp@gmail.com}
\affiliation{School of Physics, Hangzhou Normal University, Hangzhou, Zhejiang 311121, China}

\author{Lei Pan}%
\email{panlei@nankai.edu.cn}
\affiliation{School of Physics, Nankai University, Tianjin 300071, China}

\begin{abstract}

We study the thermalization dynamics of a quantum system embedded in an incommensurate potential and coupled to a Markovian thermal reservoir. The dephasing induced by the bath drives the system toward an infinite-temperature steady state, erasing all initial information-including signatures of localization. We find that initially localized states can relax to the homogeneous steady state faster than delocalized states.
Moreover, low-temperature initial states thermalize to infinite temperature more rapidly than high-temperature states---a phenomenon reminiscent of the Mpemba effect, in which hotter liquids freeze faster than colder ones. The slowest relaxation mode in the Liouvillian spectrum plays a critical role in the expedited thermalization for localized or cold initial states. Our results reveal that the combination of disordered structure and environmental dissipation may lead to non-trivial thermalization behavior, which advances both the conceptual framework of the Mpemba effect and the theoretical understanding of nonequilibrium processes in dissipative disordered systems.
\end{abstract}

\maketitle

\section{Introduction}

Nonequilibrium physics, particularly nonequilibrium dynamics, constitutes a long-standing yet perpetually vibrant research frontier that spans nearly all branches of physics. In contrast to equilibrium systems, nonequilibrium dynamics inherently involve all degrees of freedom, precluding a complete thermodynamic description via partition-function-based methods in statistics mechanics. This intrinsic complexity positions nonequilibrium dynamics among the most fundamental challenges in modern physics. Recent advances in quantum gas microscopy and controlling parameters of the system have enabled unprecedented experimental access to non-equilibrium dynamics in diverse quantum systems. Nonequilibrium dynamics have uncovered a host of exotic physical effects such as quantum many-body scarring \cite{Scar1,scar_Review1,scar_Review2,scar_Review3}, and Kardar-Parisi-Zhang universality \cite{KPZ1,KPZ2,KPZ3,KPZ4}. 

A non-intuitive relaxation phenomenon in nonequilibrium dynamics is the  Mpemba effect wherein initially warmer liquids undergo faster cooling to a target temperature than their cooler counterparts under identical external conditions. Experimental observations of the Mpemba effect span classical systems \cite{ME_classical1,ME_classical2,ME_classical3,ME_classical4,ME_classical5,ME_classical6,ME_classical7,ME_classical8,ME_classical9}. Conversely, an inverse Mpemba effect where a colder system heats up faster than a warmer one has been theoretically predicted \cite{Inverse_ME1,Inverse_ME2} and recently confirmed experimentally \cite{Inverse_ME3}. Recently, the discovery of a quantum version of Mpemba effect has been achieved through both theoretical proposals and experimental realizations \cite{QME_Exp1,QME_Exp2,QME_Exp3} in well-controlled quantum systems. The experimental realization of quantum Mpemba effects has stimulated substantial theoretical investigations into their underlying mechanisms \cite{QME1}, spanning integrable or non-integrable many-body systems \cite{QME2,QME3,QME4,QME41,QME_theory1}, many-body localizations \cite{QME5}, random circuits \cite{QME6,QME7} and diverse physical platforms \cite{QME8,QME9,QME10,QME11,QME12,QME13,QME14,QME15,QME16,QME17,QME18,QME19,QME20}. Conversely, an inverse Mpemba effect where a colder system heats up faster than a warmer one has been theoretically predicted \cite{Inverse_ME1,Inverse_ME2} and recently confirmed experimentally \cite{Inverse_ME3}.
A natural theoretical framework for describing quantum Mpemba effects (QME) or inverse QME is through open quantum systems. Quantum dissipative dynamics for open quantum systems play a pivotal role in nonequilibrium statistical physics, offering a window into emergent phenomena such as thermalization and information scrambling. This approach provides a powerful paradigm to characterize the QME by comparing thermalization pathways from initial states at different temperatures and investigating the role of decoherence in relaxation processes. 
Since the impact of the thermal bath, a quantum system with initial information such as localization properties should become delocalized completely throughout the entire system. This occurs because local dephasing mechanisms inevitably drive quantum systems toward infinite-temperature states, regardless of their Hamiltonian structure.

Localization serves as a fundamental mechanism that breaks quantum thermalization, profoundly affecting both charge transport and thermalization dynamics in quantum systems. Previous research on many-body localized systems coupled to external reservoirs has demonstrated that environmental interactions destroy localization, ultimately leading to thermal equilibrium. However, localized systems display characteristic slower relaxation, which fundamentally differs from the thermalization dynamics observed in ergodic systems. Hence, general expectations suggest extended states (closer to the steady state in configuration space) or high-temperature states should relax faster.
Motivated by recent advances in open disordered quantum systems \cite{OpenMBL1,OpenMBL2,OpenMBL3,OpenMBL4,OpenMBL5,OpenMBL6,OpenDisorder1,OpenDisorder2,OpenDisorder3,OpenDisorder4,OpenDisorder5,OpenDisorder6,OpenDisorder7,OpenDisorder8,OpenDisorder9,OpenDisorder10,OpenDisorder11,OpenDisorder12,OpenDisorder13,OpenDisorder125,OpenDisorder135,OpenDisorder14,OpenDisorder15,OpenDisorder16,OpenDisorder17,OpenDisorder18} and the experimental realization of quantum Mpemba effects \cite{QME_Exp1,QME_Exp2,QME_Exp3}, we investigate the thermalization dynamics in open quantum systems with incommensurate potentials, aiming to identify potential Mpemba-like behavior. Specifically, we examine two questions: first, whether localized initial states or extended initial states reach the maximally mixed state faster, and second, whether low-temperature initial states or high-temperature initial states are heated to infinite temperature more quickly.
%
%

\section{Theoretical Framework and Model}

A natural framework for describing the inverse Mpemba effect lies in investigating the dissipative dynamics of open quantum systems.
Consider a quantum system interacting with an environment, where the total Hamiltonian \(H_{T}\) takes the form  

\begin{align}  
	H_{T} = H_{S} + H_{E} + H_{SE},  
\end{align}  
with \(H_{S}\) and \(H_{E}\) denoting the system and environment Hamiltonians, respectively, while \(H_{SE}\) accounts for their interaction. Under the Born-Markov approximation \cite{Moy1999,Breuer2002}, tracing out the environmental degrees of freedom leads to a Lindblad master equation \cite{Lindblad1,Lindblad2} describing the system's time evolution 

\begin{align}  
	\frac{d\rho(t)}{dt} = \mathscr{L}[\rho(t)] = -i[H_{S}, \rho(t)] + \mathcal{D}[\rho(t)], \label{Lindblad_Eq}
\end{align}  
where \(\rho(t)\) is the system's density matrix, and \(\mathscr{L}\) denotes the Liouvillian superoperator which is a mathematical mapping that transforms one matrix into another with trace and positive-definition conserving. The first term governs unitary dynamics, whereas \(\mathcal{D}[\rho(t)]\) encodes dissipative effects 
\begin{align}  
	\mathcal{D}[\rho(t)] = \sum_{j} \sum_{m=1}^{M} \Gamma^{(m)}_{j} \left( O^{(m)}_{j} \rho O^{(m)\dagger}_{j} - \frac{1}{2} \left\{ O^{(m)\dagger}_{j} O^{(m)}_{j}, \rho \right\} \right). 
\end{align}  
Here $\{A, B\}\equiv AB+BA$ is the anticommutator, $O^{(m)}_{j}$ are called dissipation operator or quantum jump operator, $j$ labels lattice sites, and $M$ indicates the number of dissipation channels per site with dissipation strength \(\Gamma^{(m)}_{j}\).  
%
The open quantum system's evolution is dictated by the spectrum of superoperator \(\mathscr{L}\), with the formal solution $\rho(t)e = e^{\mathscr{L}t} [\rho_0]$. At long times, the system approaches a steady state $\rho_{ss}= \lim_{t \to \infty} \rho(t)$, corresponding to the zero-eigenvalue eigenvector of \(\mathscr{L}\). The nature of this steady state is influenced by the specific jump operators in \(\mathscr{L}\).  
The dynamics governed by the master equation admit a spectral decomposition solution
\begin{equation}
	\rho(t) = \rho_{ss} + \sum_{n=2}^{d^2} \text{Tr}({l}_n {\rho}_0) {r}_n e^{\lambda_n t}, \label{rho_extend}
\end{equation}
where $d$ denotes the Hilbert space dimension, $\rho_0$ the initial density operator, while ${r}_n$ and ${l}_n$ represent respectively the right and left eigenvectors of the Liouvillian superoperator $\mathscr{L}$ with eigenvalues $\lambda_k$, satisfying $\mathcal{L}[{r}_n] = \lambda_n {r}_n$ and $\mathscr{L}^{\dagger}[{l}_n] = \lambda_{n}^{*} {l}_n$. These eigenvalues are ordered by their real components $0 = \lambda_1 < |\text{Re}(\lambda_2)| \leq |\text{Re}(\lambda_3)| \leq \ldots\leq |\text{Re}(\lambda_{d^2})|$, all possessing non-positive real parts.

This expansion \eqref{rho_extend} reveals that the quantum dynamics decomposes into $d^2$ independent decay channels. The $\lambda_1$ channel corresponds to the steady-state solution, while all others exhibit exponential relaxation governed by $e^{\lambda_n t}$. Crucially, the decay rate scales with $|\text{Re}(\lambda_n)|$ meaning larger values indicate faster relaxation. The dominant subleading contribution comes from $\lambda_2$, representing the most persistent non-equilibrium mode. 

%

Next we introduce the system model as will discussed in this paper. We consider the system subjected to a quasi-periodic potential with system length $L$ which is described by the following Hamiltonian 
\begin{equation}
H_S= \sum_{j} V \cos(\beta j^{\alpha}) a_{j}^{\dagger} a_{j} + \sum_{j}^{L-1} J (a_{j+1}^{\dagger} a_{j} + a_{j}^{\dagger} a_{j+1}).
\end{equation}
Here, \( V \) is the potential strength, \( J \) is the hopping amplitude, \( \beta \) and \( \alpha \) are real numbers, and \( j \) is the site index. 
When $\beta$ takes rational values and $\alpha$ is integer-valued, the system reduces to the standard periodic Bloch model. For irrational $\beta$ with $\alpha \geq 2$ the system produces identical localization behavior to the random Anderson model, exhibiting complete spatial localization with identical localization lengths.
The model exhibits a mobility edge within a certain parameter regime ($0<\alpha<1$,$V<2$ with irrational $\beta$) supports mobility edges with the mobility edge at $E_c= \pm(2J-V)$ \cite{model}. 

\begin{figure*}[ht!]
	\centering
	\includegraphics[width=0.98\linewidth]{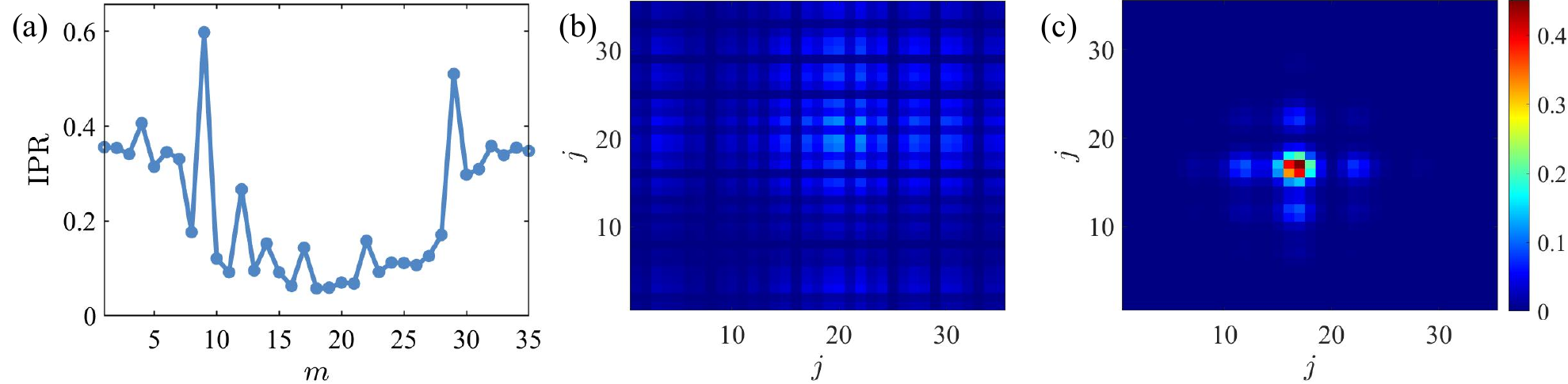}
	\caption{ Mobility Edge Display. (a) The inverse participation ratio (IPR) of the eigenstates of the Hamiltonian, where \( m \) denotes the eigenstate index ordered by increasing energy. (b) The site occupation density matrix of the localized state (indexed by \( m = 3 \)), where \( j \) denotes the site index. (c) The site occupation density matrix of the extended state (indexed by \( m = 16 \)). Here the system parameters are chosen as $V = 1.4$, $J = 1$, quasiperiodic modulation $\beta = 4\pi^2$, $\alpha = 0.7$, and the system size $L = 35$.}
	\label{fig1}
\end{figure*}
We select a specific parameter set that generates a well-defined mobility edge where the potential strength $V = 1.4$, hopping amplitude $J = 1$, quasiperiodic modulation $\beta = 4\pi^2$, and $\alpha = 0.7$, and system size $L=35$.
Figure~\ref{fig1}(a) displays the calculated inverse participation ratio (IPR), clearly revealing the coexistence of localized and extended states, with the mobility edge separating the two regimes. To further characterize these states, we examine the real-space density matrices of a representative localized state (index $m = 3$) and an extended state (index $m = 16$), which will be utilized in our later analysis. As evident from Fig.~\ref{fig1}(a), states near the spectral edges exhibit strong localization, whereas those near the band center remain extended. This distinction is further corroborated by the spatial structure of their density matrices. In Fig.~\ref{fig1}(b), the extended state ($m = 16$) displays a broad spatial distribution, with particle density delocalized across multiple lattice sites. In contrast, Fig.~\ref{fig1}(c) demonstrates that the localized state ($m = 3$) is confined to a narrow region, with negligible amplitude away from its central site. These observations align with our previous discussion on the energy-dependent localization properties. Remarkably, when studying the dissipative relaxation toward an infinite-temperature steady state, we observe that the localized state-despite initially residing at a lower effective temperature-thermalizes faster than both the high-temperature thermal state and the extended state. This counterintuitive behavior, which we attribute to the inverse QME, highlights the crucial role of localization in dictating nonequilibrium dynamics.
\begin{figure}
    \centering
    \includegraphics[width=0.95\linewidth]{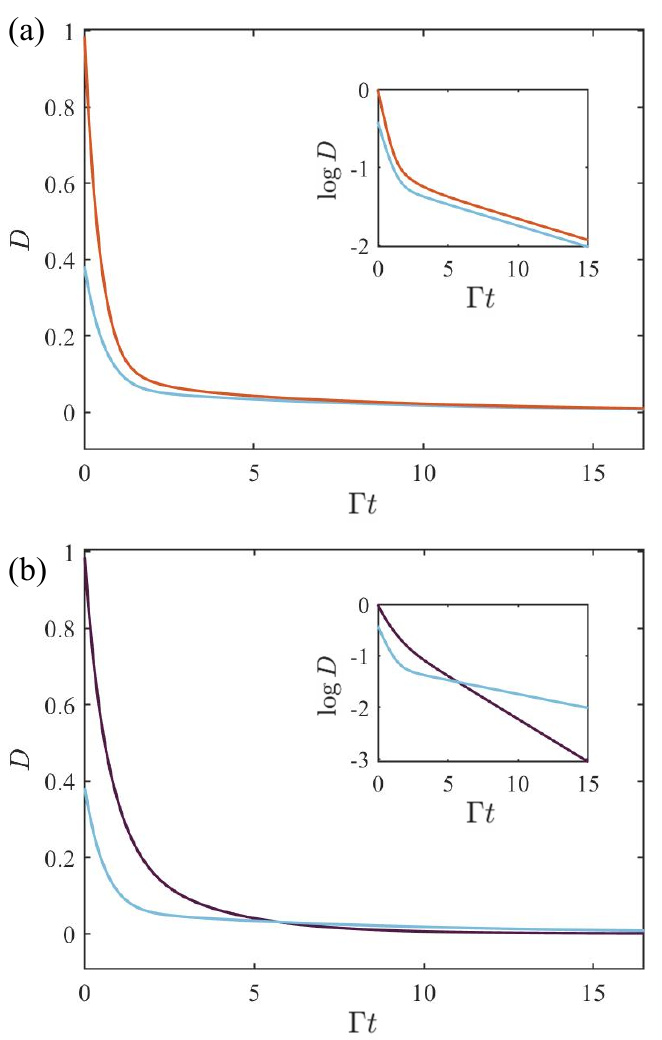}
     \caption{ The time evolution of the Frobenius distance. (a) Temporal evolution of the Frobenius distance \( D \) between the extended state (indexed by \( m = 16 \), \( T_e = 9.15\), orange curve) and the thermal state (at temperature \( T_t = 0.25\), blue curve). The upper-right inset shows the logarithmic Frobenius distance \( \ln D \). The absence of curve intersection clearly indicates no inverse QME occurs. (b) Temporal evolution of the Frobenius distance between the localized state (indexed by \( m = 3 \), \( T_l = 0.22 \), dark red curve) and the thermal state (at \( T_t = 0.25 \), blue curve). The intersection of the two curves confirms the occurrence of inverse QME.}
    \label{fig2}
\end{figure}

\begin{figure}
    \centering
    \includegraphics[width=1\linewidth]{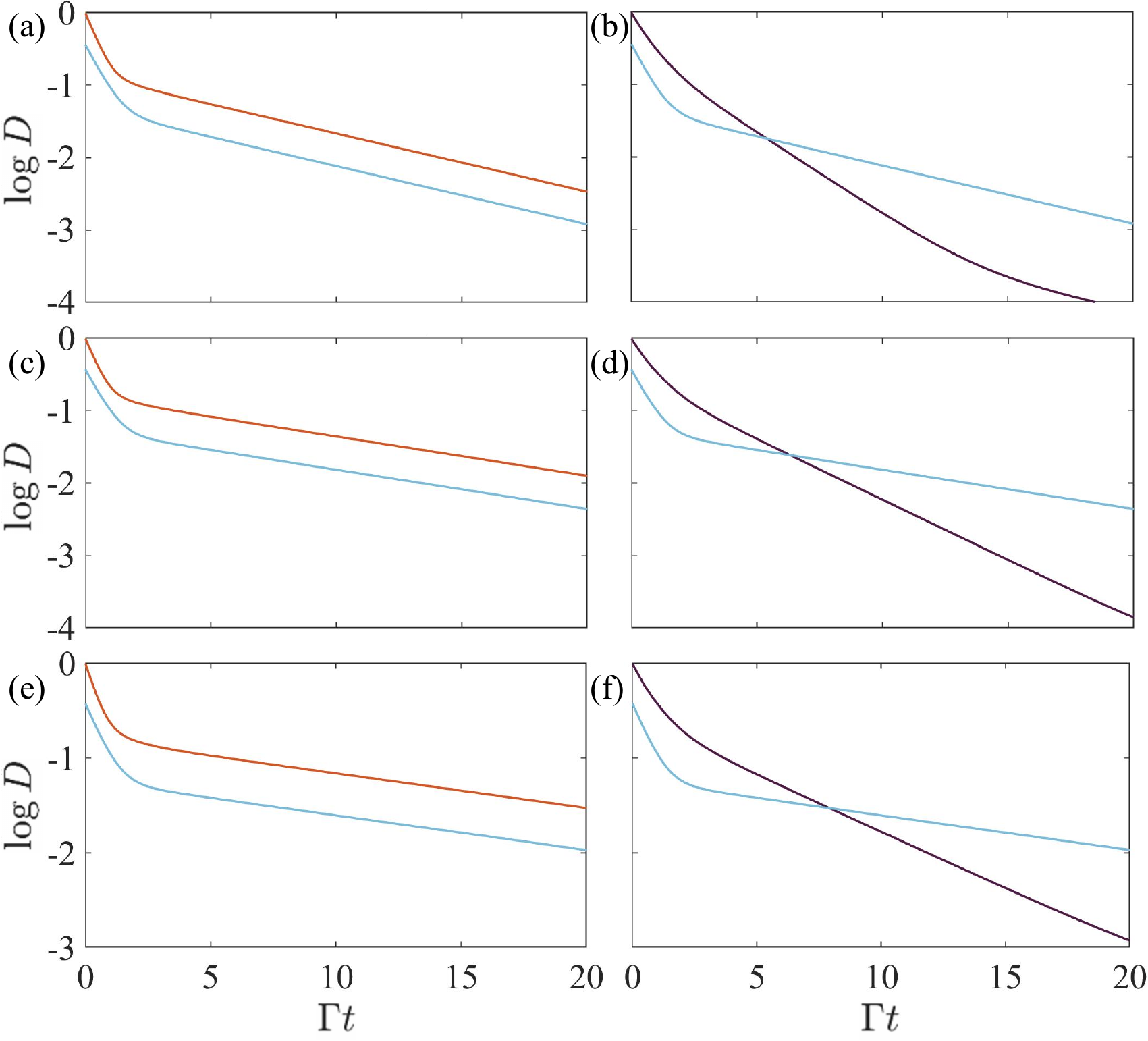}
     \caption{ Temporal evolution of \( D \) under different potential strengths \( V \). The orange, dark red, and blue curves respectively represent extended states (indexed by \( m = 16 \)), localized states (indexed by \( m = 3 \)), and thermal states at temperature \( T_t = 0.3 \), with potential strengths \( V = 1.2\) (panels (a)-(b)), \( V = 1.4 \) (panels (c)-(d)), and \( V =1.6 \) (panels (e)-(f)), where in all cases the temperatures satisfy \( T_{l} < T_t < T_{e} \), with \( T_{l} \), \( T_{t} \), \( T_{e} \) and  denoting the localized, thermal and extended state temperatures respectively. }    	
    \label{fig3}
\end{figure}

\begin{figure}
    \centering
    \includegraphics[width=1\linewidth]{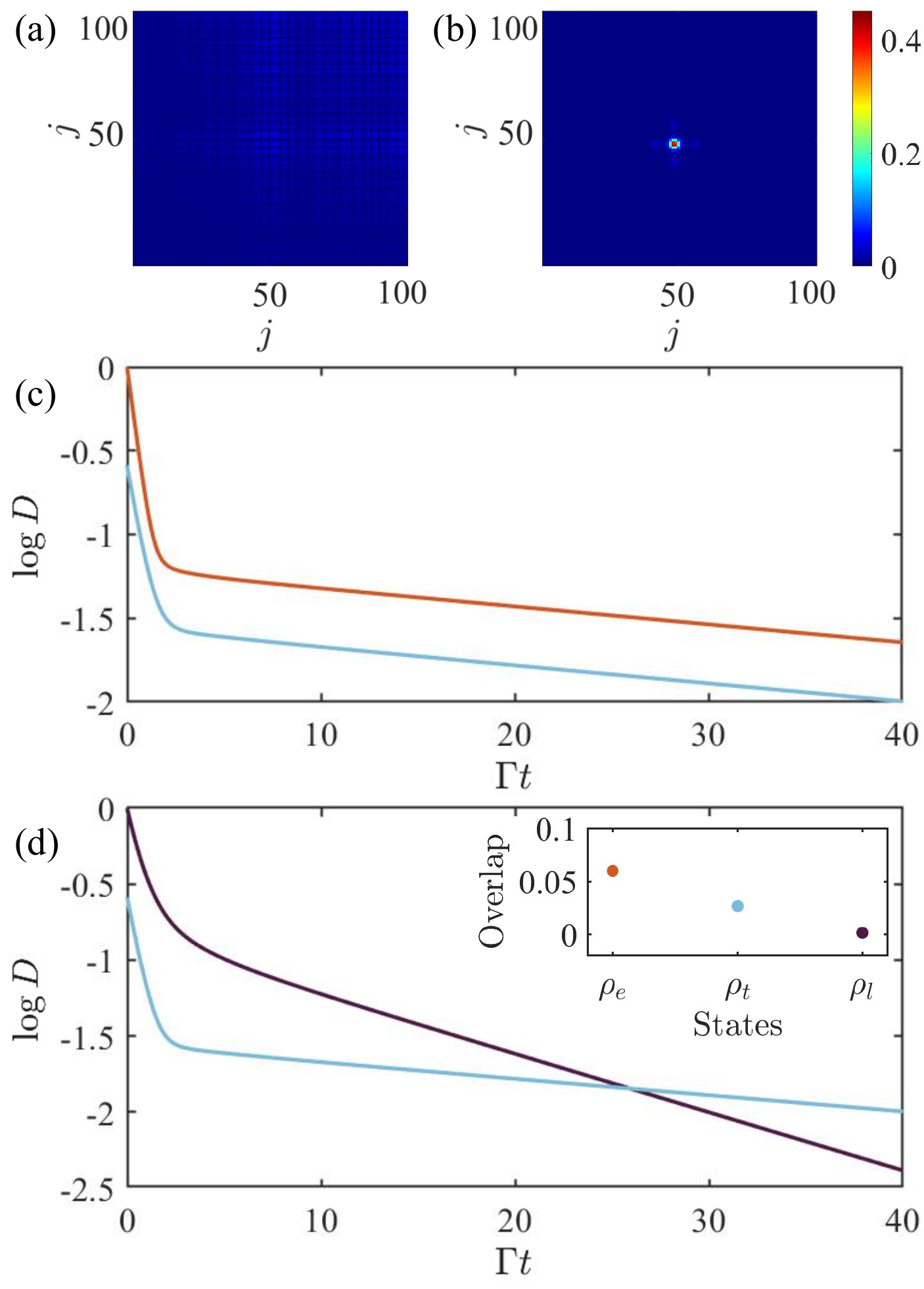}
    \caption{
   The inverse QME for a lattice with \( L = 100 \) sites. 
   (a) Real-space density matrix of the extended state (energy index $m = 47$); 
   (b) Real-space density matrix of the localized state (energy index $m = 7$); 
   (c) Temporal evolution of the Frobenius distance between the extended state (orange curve) and thermal state (blue curve); 
   (d) Temporal evolution between the localized state (dark red
   curve) and thermal state. 
   The initial state temperatures are $T_l = 0.23$, $T_t = 0.3$, and $T_e =12.26$ for the localized, thermal, and extended states respectively. 
   The inset in (d) displays the absolute value of the overlap between initial states and the slowest mode. }
    \label{fig4}
\end{figure}

\section{Expedited thermalization and inverse quantum Mpemba effect}

In this section, we investigate heating dynamics induced by quantum dissipation, revealing the inverse QME in such systems. As established previously, the dissipative evolution of the system is governed by the Lindblad master equation \eqref{Lindblad_Eq}. For simplicity, we consider the case where each lattice site couples to a single dissipation channel ($M=1$), with uniform dissipation strength $\Gamma$ across all sites. We choose the dissipative operator as dephasing, i.e., $O_j = n_j = a_j^{\dagger} a_j$ for all $j$.
The dissipative operator is Hermitian, which ensures that the steady state of the system is the maximally mixed state, also known as the infinite-temperature thermal state (see appendix \ref{App1}). In any basis representation, its steady-state density matrix is  
\begin{equation}
\rho_{\text{ss}} = \frac{1}{L} {\bf I}
\end{equation}
Here, \( {\bf I}\) is the identity matrix, whose dimension is equal to the dimension of the system's Hilbert space. Under the influence of the above dissipation, the system will be driven to the steady-state \( \rho_{\text{ss}} \) regardless of its initial state.
To quantify the inverse QME, we introduce the Frobenius norm as a measure of the distance between a given state $\rho$ and the steady state $\rho_{\text{ss}}$: 
\begin{equation} 
	\mathcal{D}(\rho) = \|\rho - \rho_{\text{ss}}\|_F = \sqrt{\text{Tr}[(\rho - \rho_{\text{ss}})^\dagger (\rho - \rho_{\text{ss}})]}. 
	\end{equation} 
The inverse QME manifests when a state initially farther from equilibrium (larger $D$) relaxes faster than a closer state.
 
We analyze the time evolution of both the extended (Fig.~\ref{fig1}(b)) and localized (Fig.~\ref{fig1}(c)) states, comparing their relaxation dynamics with a thermal state at temperature $T_t$. The thermal state is defined as 
\begin{equation}
	\rho_{T_t} = \frac{1}{\mathcal{Z}}\sum_n e^{-E_n/T_t}|n\rangle\langle n|, \quad \mathcal{Z} = \sum_n e^{-E_n/T_t},
\end{equation}
where $\{|n\rangle\}$ are Hamiltonian eigenstates with eigenvalues $\{E_n\}$ and \( T_t \) is the temperature of thermal state. 
\par

Figure~\ref{fig2} plots distinct relaxation behaviors for different initial states. The extended state maintains a larger Frobenius distance $\mathcal{D}$ from the steady-state compared to the thermal state at a temperature $T_t$ throughout the entire evolution, both at initial time ($t=0$) and in the long-time limit ($t\to\infty$). This persistent separation indicates the absence of inverse QME between these states. More remarkably, while the localized state initially shows greater deviation from equilibrium ($\mathcal{D}_l(0) > \mathcal{D}_t(0)$), its relaxation trajectory crosses that of the thermal state at finite time, ultimately reaching smaller $\mathcal{D}$ values in the long-time regime. This characteristic crossing demonstrates a clear signature of the inverse QME, where the initially more distant localized state thermalizes faster than both the thermal reference state and the extended state. The accelerated relaxation of the localized state occurs despite its greater initial distance from equilibrium, challenging conventional expectations that states farther from equilibrium should take longer to relax.

The temperature analysis further corroborates the existence of the inverse QME in our system. The three initial states exhibit distinct effective temperatures: $T_l $ for the localized state, $T_t$ for the thermal state, and $T_e$ for the extended state, establishing the clear hierarchy $T_l < T_t < T_e \ll T_{\text{ss}} = \infty$. Remarkably, the coldest localized state ($T_l$) demonstrates the fastest relaxation to the infinite-temperature steady state, violating conventional thermal relaxation expectations. This manifests as two concurrent effects: the localized state thermalizes faster than both the intermediate-temperature thermal state and the much hotter extended state. The persistence of this temperature-inverted relaxation behavior highlights the non-thermal nature of inverse QME in our system. The localized state's accelerated dynamics cannot be explained by temperature considerations alone, implying that quantum disordered systems can exhibit fundamentally different thermalization mechanisms that transcend simple temperature-based predictions.

To reveal the physical mechanism of the inverse QME, we investigate the asymptotic behavior that emerges from \eqref{rho_extend} in the long-time limit 
\begin{equation}
	\rho(t) \approx \rho_{ss} + e^{\text{Re}(\lambda_2)t} \left[ \text{Tr} \left( l_2 \rho_0 \right) r_2 e^{i \text{Im}(\lambda_2)t} \right],
\end{equation}
meaning the relaxation speed depends critically on the projection $\text{Tr}(l_2\rho_0)$ between the initial state and the slowest-decaying mode $r_2$. Smaller overlap values lead to faster equilibration, as the system requires less time to relax to the steady-state. This is crucial for understanding the mechanisms behind accelerated thermalization and the inverse QME.
Our analysis demonstrates that localized initial states exhibit markedly weaker overlap with $r_2$ compared to their extended counterparts, explaining their accelerated approach to steady-state despite appearing more distant in certain physical metrics.

It is worth to emphasis that the inverse QME demonstrated in our system exhibits remarkable robustness against parameter variations. As shown in Fig.~\ref{fig3}, the characteristic relaxation behavior persists across a finite range of potential strengths $V$, with the localized state consistently showing accelerated thermalization compared to both thermal and extended states. This parametric stability suggests that the observed phenomenon is not fine-tuned but rather a generic feature of the system's dissipative dynamics.

Furthermore, the effect maintains its distinctive signature in larger systems. Numerical simulations for a lattice of $L=100$ sites (Fig.~\ref{fig4}) confirm that the essential physics remains unchanged: the localized state at lower effective temperature continues to thermalize faster than higher-temperature states. This can be clearly understood via overlaps between distinct initial states and the slowest-decaying mode as plotted in the inset of Fig. \ref{fig4} (d). The extended state exhibits the maximum overlap with the slowest mode, while the localized state shows nearly vanishing overlap which provides a physically justified explanation for the emergence of inverse QME. The system-size independence demonstrates that the QME is not merely a finite-size artifact but persists in the thermodynamic limit, reinforcing its significance as a genuine non-equilibrium quantum phenomenon.

The combination of parameter robustness and scalability with system size provides strong evidence that the observed inverse QME represents a fundamental characteristic of dissipative quantum systems with localized states, rather than a special case limited to specific parameters or small systems.
 

\section{Conclusion and outlook}
In this work, we have systematically investigated the thermalization dynamics of quantum systems embedded in incommensurate potentials and coupled to Markovian reservoirs, uncovering a robust inverse QME phenomenon where initially colder and more localized states thermalize faster to infinite temperature than their warmer and extended counterparts. Through a combination of analytical and numerical approaches, we have demonstrated that this counterintuitive behavior stems from the reduced overlap between localized states and the slowest-decaying Liouvillian modes, providing a clear physical mechanism for the accelerated thermalization. The effect persists across parameter variations and system sizes, indicating its fundamental nature in dissipative quantum systems with localization properties.

Our findings advance the understanding of nonequilibrium dynamics and establish a connection between quantum localization and anomalous thermalization phenomena. They demonstrate that temperature hierarchy alone cannot predict relaxation timescales in quantum systems.

Our work can be directly extend to other localized systems with different disorder potentials which can be experimentally realized in cold atom systems using existing techniques for engineering disorder or quasiperiodic potentials and controlled dissipation \cite{Exp1,Exp2,Exp3,Exp4,Exp5,Exp6,Exp7,Exp8,Exp_new1,Exp_new2,Exp_new3}.

The robustness of inverse QME suggests it may be observable in various quantum simulation platforms \cite{simulation,simulation1,simulation2}, potentially opening new avenues for controlling quantum thermalization processes. Future theoretical work could fruitfully explore the relationship between our findings and other manifestations of the QME across different physical systems, as well as investigate the role of different types of system-environment couplings.

\section*{Acknowledgements}
The work is supported by the National Natural Science Foundation of China (Grant No. 12304290) and the Fundamental Research Funds for the Central Universities. \\

\appendix
\section{Proof of maximally mixed steady-state under hermitian dissipation operators} \label{App1}
 
In this appendix, we prove the conclusion that a steady-state is maximally mixed state when dissipation operators are hermitian. For simplicity, we assume the dissipation operator is local dephasing, namely, $O_{j}= n_{j}$ and consider a generic tight-binding model.  Defining the matrix element $\rho_{i, j}(t)=\langle { i}| \rho(t)|{ j}\rangle$ in the local basis, where $| j\rangle=\hat{c}_{ j}^{\dagger}|0\rangle$ is the single-particle state at the lattice site $ j$, we can obtain the time evolution equations from the Lindblad master equation 
	\begin{align}
		\frac{\mathrm{d}}{\mathrm{~d} t} \rho_{i, j}= & -i J\left(\rho_{i, j+1}+\rho_{i, j-1}-\rho_{i+1, j}-\rho_{i-1, j}\right) \nonumber \\
		& +i\left(\epsilon_{\it j}-\epsilon_{\it i}\right) \rho_{\it {i, j}}-\Gamma \rho_{i, j}\left(1-\delta_{i, j}\right), \label{Diff_Eq}
	\end{align}
	where $\epsilon_{\it j}$ ($\epsilon_{\it i}$) is the onsite potential at the site $j$ ($i$), $J$ is the hopping strength, and $\delta_{i, j}$ denotes the (Kronecker) delta function. We can prove that the steady-state solution $\rho_{\rm ss}$ to above equation which corresponds to $\frac{\mathrm{d}}{\mathrm{~d} t} \rho_{\it {i, j}}=0$ for any matrix element, is the maximally-mixed state, namely the steady-state density matrix $\rho_{\rm ss}$ is proportional to the identity matrix
	
	\begin{align}
	\rho_{\rm ss}=\frac{1}{L}{\bf I}.
	\end{align}
	
	First, terms $\Gamma \rho_{\boldsymbol{i, j}}\left(1-\delta_{\boldsymbol{i, j}}\right) $ in the differential equation \eqref{Diff_Eq} lead to all off-diagonal terms of density matrix $\rho_{i, j}$ (${ i}\neq{ j}$ ) decay to zero in the steady state, leaving only the diagonal elements.
	Second, the diagonal elements of steady-state density matrix are uniform due to symmetry. This is because if the diagonal elements are not uniform, the probability distribution on lattice sites would flow from higher-population to lower-population sites due to the symmetric hopping terms in the Hamiltonian, which is in contradiction with the steady-state condition.

\end{document}